\documentclass[final,1p,times]{elsarticle}

\usepackage{graphicx}

\usepackage{bbm, bm, mathrsfs, amsthm, amsmath, amssymb, amsfonts, mathrsfs}

\usepackage{color}

\usepackage[bookmarks = true, citecolor=blue, colorlinks = true, linkcolor=magenta, urlcolor = blue]{hyperref}

\journal{\textsf{\ Annals of Physics\ {\bf 353} 326--339 (2015)}}

\begin{document}

\begin{frontmatter}



\title{Classical systems can be contextual too:\\Analogue of the Mermin-Peres square}


\author{Pawel Blasiak}

\address{Institute of Nuclear Physics, Polish Academy of Sciences\\
ul.~Radzikowskiego~152, PL 31-342 Krak\'ow, Poland}
\ead{Pawel.Blasiak@ifj.edu.pl}

\begin{abstract}
Contextuality lays at the heart of quantum mechanics. In the prevailing opinion it is considered as a signature of "quantumness" that classical theories lack. However, this assertion is only partially justified. Although contextuality is certainly true of quantum mechanics, it cannot be  taken by itself as discriminating against classical theories. Here we consider a representative example of contextual behaviour, the so-called Mermin-Peres square, and present a discrete toy model of a bipartite system which reproduces the pattern of quantum predictions that leads to contradiction with the assumption of non-contextuality. This illustrates that quantum-like contextual effects have their analogues within classical models with epistemic constraints such as limited information gain and measurement disturbance.
\end{abstract}


\end{frontmatter}

\section{Introduction}
\label{Introduction}

Weirdness of quantum mechanics is usually presented by way of contradiction with simple classical intuitions that we hold about the world. Many of these paradoxes are often raised in the debates on the interpretation of the theory, and in particular the possibility of its hidden variable account. There are two major no-go theorems to the effect that such a description is subject to severe constraints: the Bell's theorem shows violation of locality~\cite{Be66,Be93}, and the Kochen-Specker theorem contradicts the premise of non-contextuality~\cite{Sp60,KoSp67,Be66,Be93}. Clearly, this presents a challenge to our intuitive understanding of quantum theory and sets the bar high for hidden variable models~\cite{Pe95,Me93}. In this work, we are concerned with quantum contextuality which, in a nutshell, says that there is no consistent assignment of values to quantum mechanical observables and whose objective existence is independent of the context of other observables that are being simultaneously measured. This surprising result does not quite fit in with our naive conception of the act of measurement revealing the value of an observable irrespective of whether or not other quantities are also being observed. Are we then bound to draw the conclusion after Asher Peres that "\emph{unperformed experiments have no results}"~\cite{Pe78}? If so, how to answer the David Mermin's dramatic question: "\emph{Is the moon there when nobody looks?}"~\cite{Me85}. It is not clear what is a good way out of this conundrum and, in particular, what form acceptable hidden variable models could take to that effect. Certainly, at this point one should seriously reflect on the John Bell's dictum ''\emph{... what is proved by impossibility proofs is lack of imagination}''~\cite{Be82}. It might suggest that perhaps revision of the relation between the concept of observable and measurement is required for better understanding of the theory. In this article we attempt to explore this possibility and show that careful distinction between these concepts opens a way for methodological discussion of contextual effects in classical systems too.

Difficulty in making sense of contextuality in classical terms often prompts to consider it as a signature distinguishing between quantum and classical realms. Indeed, the possibility of contextual hidden variable models aiming at reconstruction of quantum predictions is hardly explored. On the other hand, the hypothesis of non-contextual hidden variable models has been thoroughly investigated~\cite{Pe95} and proved to be directly testable~\cite{Ca08d,ThKuLeSoKa13}. In particular, many state-independent quantum-contextuality experiments have been recently performed e.g. with trapped ions~\cite{KiZaGeKlGuCaBl09}, photons~\cite{AmRaBoCa09,DAHeAmNaBoScCa13} and magnetic-resonance systems~\cite{MoRyCoLa10}. Essentially all of them boil down to checking of the pattern captured in the so called Mermin-Peres square~\cite{Me90,Pe90}. Certainly, these results provide compelling evidence for contextual behaviour in these experimental setups, thus pushing the project of hidden variable models to the less explored contextual camp.
In this work we present a simple probabilistic model which reproduces the pattern of quantum observables considered in the Mermin-Peres square and demonstrates quantum-like contextuality in a classical bipartite system. One immediate feature of the presented model is state-independent violation of contextuality inequality~\cite{Ca08d} in accord with quantum mechanical predictions.

Many results suggest that quantum states can be understood as states of knowledge. Strong evidence in favour of this view is given, in particular, by concrete models providing classical analogues of various phenomena typically associated with strictly quantum mechanical characteristics~\cite{Sp07,Ha99,Ki03,La09a,DaPlPl02,KlGuPoLaCa11,BaRuSp12,La12,SzKlGu13,Bl13}. Most notable in this respect is the Spekkens' toy model~\cite{Sp07} reproducing a surprisingly large array of effects in a simple discrete system. This work has recently sparked a lot of interest and hope for $\psi$-\emph{epistemic} reconstructions of quantum theory, in which quantum state is essentially understood as a state of knowledge about some \emph{ontic} reality subject to \emph{epistemic} restrictions~\cite{HaSp10} (see also~\cite{Ha04,Sp05,HaRu07,Mo08,WaBa12,PuBaRu12} for discussion of various properties and structural constraints to be satisfied by such reconstructions). However, we should note that none of these models treats the problem of contextuality or non-locality in a straightforward and uncontrived manner. These features hold the stage presenting a great challenge to the proponents of the epistemic view. In this paper, we discuss a simple classical model which demonstrates that contextual behaviour can be understood as an effect of limited information gain and post-measurement state disturbance. A distinctive feature of the model is that it is based on a straightforward picture of a \emph{bipartite} system being composed of two separate components whose behaviour is dictated solely by classical correlations (cf. different ontologies proposed in~\cite{La09a,KlGuPoLaCa11,La12,SzKlGu13,LeJeBaRu12,EmDeSuVe13}). Moreover, it will be shown to reconstruct all aspects relevant to quantum contextuality captured in the Mermin-Peres square.

The paper is organised as follows. We begin with a brief account of the Mermin-Peres square and the problem of contextuality in quantum theory. Then we proceed to the construction of its classical analogue by first considering elementary systems, defining local measurement procedures and discussing the notion of observable from the operational point of view. In the next step, we extend this framework to the bipartite system and introduce additional non-local measurement procedure which completes the construction. We conclude by explaining how contextuality comes about in the model and show that observables in the model properly reconstruct the pattern captured in the Mermin-Peres square, thus providing a classical analogue thereof. 

\section{The Mermin-Peres square}
\label{MP-square}

\begin{figure}[t]
\begin{center}
\includegraphics[width=0.6\textwidth]{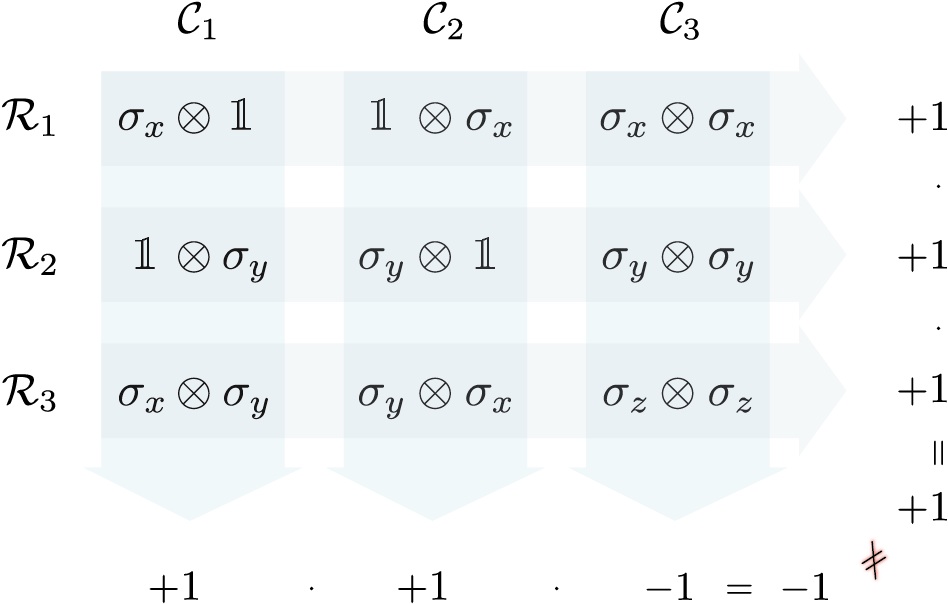}
\end{center}\caption{\label{Quantum-Mermin-Peres-Square} \textbf{The Mermin-Peres square.} Demonstration of fundamental inconsistency in assignment of non-contextual values to nine observables in a bipartite system of two qubits. Arrangement in the square is such that: (i) three observables along each row ($\mathcal{R}$) and column ($\mathcal{C}$) commute between each other, and (ii) their product is equal to $\mathbbm{1}$ except for column $\mathcal{C}_3$ where the product is $-\mathbbm{1}$. Kochen-Specker type argument assumes that these observables have preassigned values $\pm1$ which do not depend on the measurement context and for simultaneously measurable (i.e. commuting) observables algebraic relations are preserved. Together with (i) and (ii) it implies that row and column wise products of values assigned to the respective observables are as indicated to the right and bottom of the square. However, this leads to inconsistency since the product of all nine values gives contradictory results $\pm1$ depending on how the terms are grouped (i.e. multiplying first by rows one gets $+1$, while grouping terms according to columns first gives $-1$).}
\end{figure}

Notorious complexity of the original Kochen-Specker proof~\cite{KoSp67} led to substantial refinements of the argument which, now, can be presented in an intuitively accessible and instructive way~\cite{Pe95,Me93}.
Arguably, the simplest demonstration of contextuality in quantum theory is the so-called Mermin-Peres square which establishes the result in four dimensions~\cite{Me90,Pe90}. In this example, to reach a contradiction with non-contextuality one considers nine observables on a system of two spin-$\tfrac{1}{2}$ particles (qubits) arranged in a table as shown in Fig.~\ref{Quantum-Mermin-Peres-Square}.
Here, $\sigma_{x}$, $\sigma_{y}$ and $\sigma_{z}$ are spin observables along $\hat{x}$, $\hat{y}$ and $\hat{z}$ axes with possible outcomes (eigenvalues) $\pm1$. The arrangement in Fig.~\ref{Quantum-Mermin-Peres-Square} is such that observables in each row and column are compatible, i.e. commute between each other, and hence can be \emph{simultaneously} measured. However, it is crucial to observe that all six measurements $\mathcal{R}_1$, $\mathcal{R}_2$, $\mathcal{R}_3$, $\mathcal{C}_1$, $\mathcal{C}_2$ and  $\mathcal{C}_3$ require different apparatus setups (arranged in a sequential measurement~\cite{KiZaGeKlGuCaBl09,AmRaBoCa09,DAHeAmNaBoScCa13,MoRyCoLa10}) which determine the so-called \emph{context} in which a given observable is measured. For example, measurement of the phase bit $\sigma_x\otimes\sigma_x$ can be realised either in the local procedure $\mathcal{R}_1$ (by multiplying out outcomes of $\sigma_x$ measurements on individual particles) that refers to the basis $|x\pm\rangle\otimes|x\pm\rangle$, or alternatively via $\mathcal{C}_3$ being the non-local measurement in the Bell basis $|\phi^\pm\rangle$, $|\psi^\pm\rangle$. In a similar manner, one identifies measurements $\mathcal{R}_1$, $\mathcal{R}_2$, $\mathcal{C}_1$, $\mathcal{C}_2$ as \emph{local} procedures, while $\mathcal{C}_3$ and $\mathcal{R}_3$ as \emph{non-local} ones.

The problem of contextuality addresses the question whether it is possible to think of an observable as having a definite (but unknown) value before it is measured and being revealed only in experiment. A theory is said to be \emph{non-contextual} if such an assignment of values to all observables is possible irrespective of the context of other observables that are simultaneously measured (i.e. details of the experimental setup). For example, in the case of the Mermin-Peres square this would mean that measurement procedures $\mathcal{R}_1$ and $\mathcal{C}_3$ should reveal the same value of the phase bit $\sigma_x\otimes\sigma_x$. In general, if quantum theory were non-contextual then it would admit assignment of values $\pm1$ to all observables in Fig.~\ref{Quantum-Mermin-Peres-Square}. However, this is not possible if the assigned values are to preserve algebraic relations between simultaneously measurable (i.e. commuting) observables. To reach the contradiction in the Mermin-Peres square it is enough to observe that the product of observables in each row and column is equal $\mathbbm{1}$ except for column $\mathcal{C}_3$, where the product is $-\mathbbm{1}$. Accordingly, the assignment of values in Fig.~\ref{Quantum-Mermin-Peres-Square} should have the same property, i.e. the values multiplied out along rows and columns give $+1$ (the only eigenvalue of $+\mathbbm{1}$) except for the last column $\mathcal{C}_3$ where the product is equal to $-1$ (the only eigenvalue of $-\mathbbm{1}$). This leads to logical inconsistency since the product of all nine values in the square calculated row and column wise respectively give conflicting results $+1$ and $-1$.
As a consequence, one is bound to conclude that quantum theory is \emph{contextual}. This means that, the value of a quantum-mechanical observable nontrivially depends on what commuting set it is actually being measured with (i.e. what is the experimental setup).

Contradiction attained in the Mermin-Peres square rests upon the assumption of the so-called \emph{counterfactual definiteness}, which ascribes values (or properties) to observables independent of whether or not the measurement actually takes place. It rules out the possibility of reconstructing quantum-mechanical predictions in terms of non-contextual hidden variables, but remains salient about less explored contextual models that are immunised against arguments of the Kochen-Specker type.

\section{Classical analogue of the Mermin-Peres square}
\label{Results}

Having discussed the Mermin-Peres square, and how it shows contextuality of quantum observables, we proceed to the main result of this paper which is demonstration of a purely classical system that exhibits just the same behaviour.
More specifically, we will describe a bipartite system and measurement procedures that will follow the pattern of \mbox{(non-)commutativity} and \mbox{(non-)locality} captured in Fig.~\ref{Quantum-Mermin-Peres-Square}. In other words, the model will show the same character of contextuality as the quantum case described in the Mermin-Peres square.

From now on we switch to the classical realm and the conventional probabilistic setup where a system is described by probabilities (epistemic states) on a well specified sample space of ontic states (hidden variables)~\cite{Sp07,HaSp10} . We begin with the definition of a bipartite system as composed of two separable components. Then, we explicitly describe local and non-local measurement procedures paying special attention to post-measurement disturbance. This will allow for discussion of sequential measurements, compatible observables, and the meaning of measurement context. Furthermore, we will demonstrate that the structure of so defined (classical) observables coincides with the Mermin-Peres square.

\subsection{Definition of the system} 

\begin{figure}[t]
\begin{center}
\includegraphics[width=0.75\textwidth]{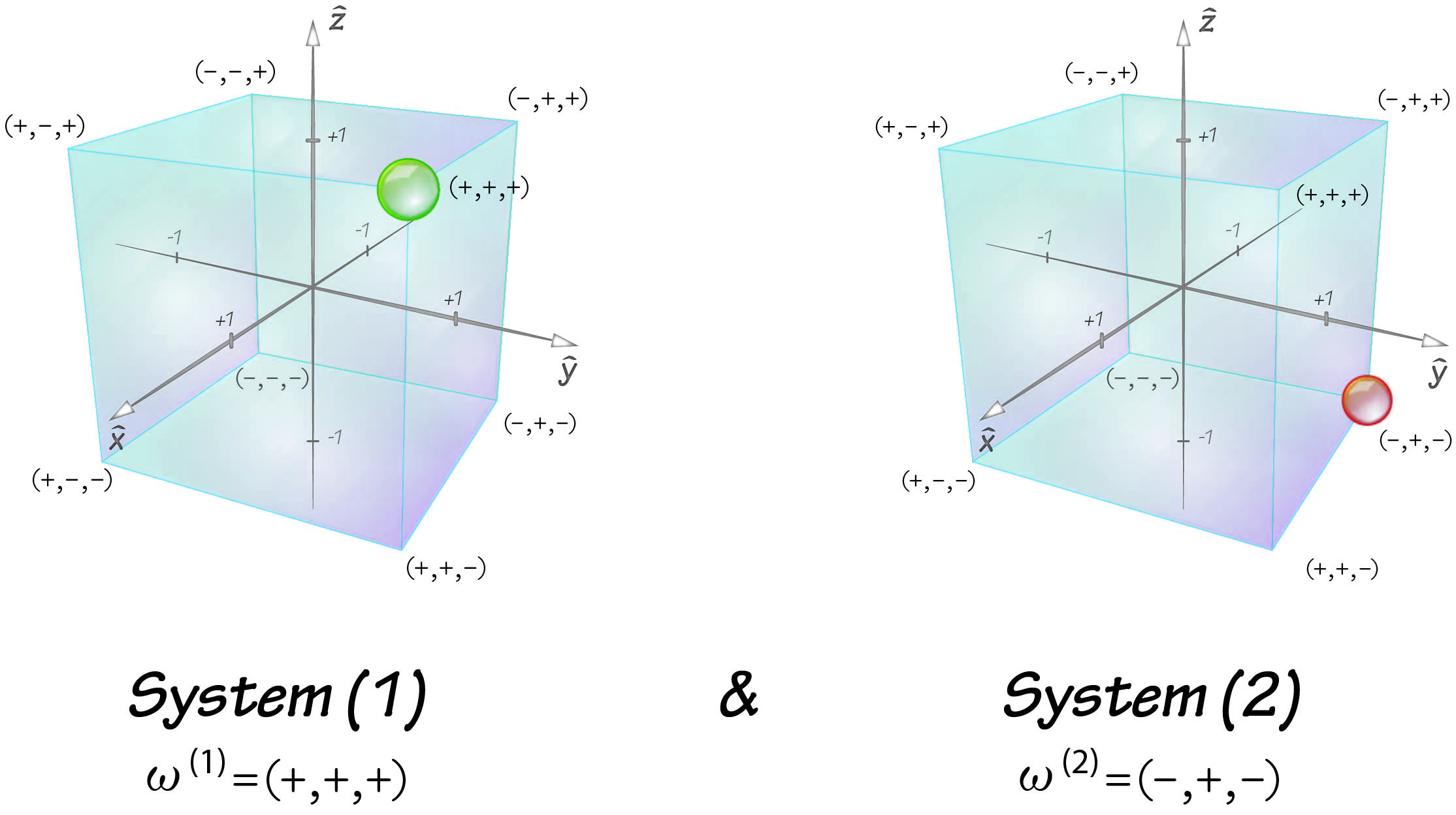}
\end{center}
\caption{\label{Fig1}
\textbf{Bipartite system.} Vertices of the cubes, labelled by cartesian coordinates $\omega=(x,y,z)$ with $x,y,z=\pm1$, represent (ontic) states of the elementary systems. Bipartite system consists of two components, called system (1) and system (2), and the joint ontic state is given by $(\omega^{(1)},\omega^{(2)})$, where $\omega^{(i)}=(x_i,y_i,z_i)$ are states corresponding to the individual components ($i=1,2$).}
\end{figure}

Let us imagine \emph{elementary system} as a cube whose vertices represent \emph{ontic states}, i.e. the sample space consists of eight states $\Omega=\left\{\omega_1,...,\omega_8\right\}$. For future convenience, we will think of the cube as fixed in the cartesian reference frame with vertices at points $x,y,z=\pm1$ and 
adopt the convention in which states are labelled by triples $\omega=(x,y,z)$, see Fig.~\ref{Fig1}. Standard description of such an elementary system consists in specifying its \emph{epistemic state} which is a probability distribution over (ontic) states in $\Omega$, i.e. it is described by a vector $\bm{p}$ in the probability simplex $\Delta=\{(p_1,...,p_8):{\sum}_{\mu=1}^8 p_\mu=1,\ p_\mu\geqslant0\}$. Note that $\bm{p}=\sum_{\mu}\,p_{\mu}\,\boldsymbol\delta_\mu$, where $\boldsymbol\delta_\mu$ is the "basis" of extremal epistemic states corresponding to the system being definitely in the ontic state $\omega_\mu$.

Now, we are in position to define a \emph{bipartite system} as composed of two elementary ones called system $(1)$ and system $(2)$. Then the joint sample space is the cartesian product $\Omega^{(1\,\&\,2)}\equiv\Omega^{(1)}\times\,\Omega^{(2)}$. It can be represented by two cubes with the joint ontic state $(\omega^{(1)},\omega^{(2)})$ being completely specified by the ontic states $\omega^{(i)}=(x_i,y_i,z_i)$ of the individual elementary systems, see Fig.~\ref{Fig1}. Note that it is quite appropriate to think of them as two distinct classical particles with internal degrees of freedom that can be (spatially) separated. Epistemic state of the system is described by a vector $\bm{p}$ in the probability simplex $\Delta^{(1\,\&\,2)}\equiv\Delta^{(1)}\otimes\Delta^{(2)}$. In other words,
$\bm{p}=\sum_{\mu,\nu}\,p_{\mu\nu}\ \boldsymbol\delta_\mu\otimes\,\boldsymbol\delta_\nu$ is a probability distribution ($p_{\mu\nu}\geqslant0$ and ${\sum}_{\mu,\nu}\,p_{\mu\nu}=1$) over the ontic states in $\Omega^{(1\,\&\,2)}$, and the distribution $\boldsymbol\delta_\mu\otimes\,\boldsymbol\delta_\nu$ corresponds to both systems being definitely in the respective ontic states ${\omega_\mu}^{(1)}$ and ${\omega_\nu}^{(2)}$. Note that from the construction such a probabilistic account allows only for classical correlations between elementary systems.

\subsection{Measurements}

The concept of measurement is subtle as its role is two-fold. First of all, it reveals information about the system, and secondly it effects change leaving the system in some post-measurement state. Complete description of the measurement procedure should give account of both these aspects. Although peripheral for a single-shot experiment, post-measurement state disturbance is crucial in the analysis of sequential measurements. As we will see, it will turn out responsible for quantum-like behaviour of the probabilistic model that we discuss below.

\subsubsection{Local measurements}

\begin{figure}[t]
\begin{center}
\includegraphics[width=0.85\textwidth]{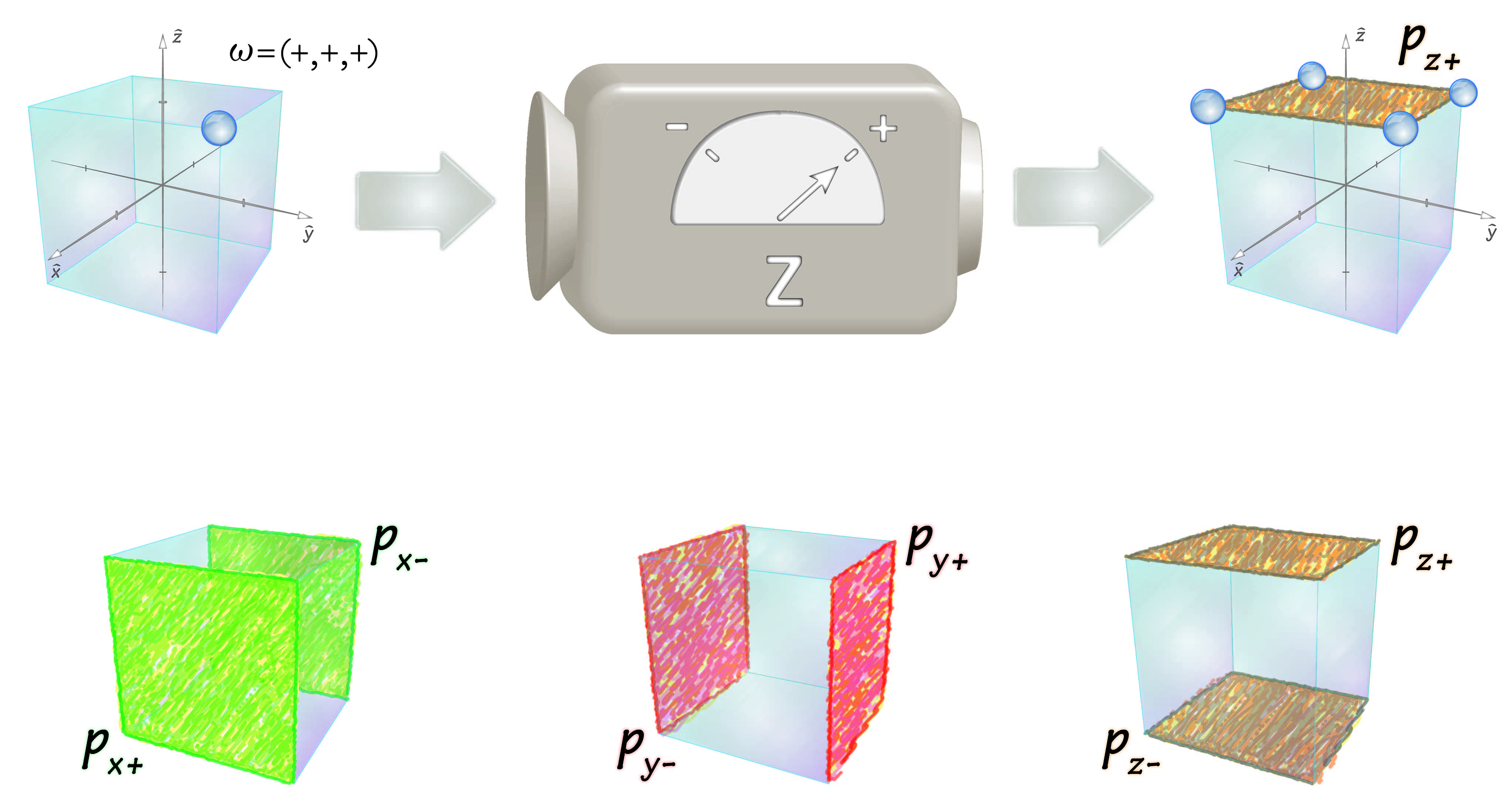}
\end{center}\caption{\label{Fig2} \textbf{Prototype of local measurement on the elementary system.} At the top, measurement $\textbf{Z}$ distinguishes between the upper ($z=+1$) and the lower ($z=-1$) face of the cube, and depending on the result leaves the system in state $\bm{p}_{z+}$ or $\bm{p}_{z-}$. At the bottom, illustration of probabilistic states $\bm{p}_{x\pm}$, $\bm{p}_{y\pm}$ and $\bm{p}_{z\pm}$ produced in the respective measurements $\textbf{X}$, $\textbf{Y}$ and $\textbf{Z}$.}
\end{figure}

We begin by defining three kinds of measurements on the elementary system which tell on which of the chosen pair of opposite faces of the cube the system resides, i.e.:\vspace{0.1cm}

\begin{tabular}{lclcl}
Measurement\ \ $\mathbf{X}$&:&front ($x=+1$)&vs.&back ($x=-1$)\ ,\vspace{0.1cm}\\
Measurement\ \ $\mathbf{Y}$&:&right ($y=+1$)&vs.&left ($y=-1$)\ ,\vspace{0.1cm}\\
Measurement\ \ $\mathbf{Z}$&:&up ($z=+1$)&vs.&down ($z=-1$)\ .\vspace{0.1cm}
\end{tabular}

\noindent In other words, the measurement reveals only \emph{one bit} of information about the chosen coordinate of the system's ontic state $\omega=(x,y,z)$. To complete the definition we assume that after the measurement the system is left with equal probability in one of the four ontic states in $\Omega$ that are compatible with the result. See Fig.~\ref{Fig2} for schematic illustration. More precisely, for measurement $\mathbf{X}$ depending on the result $x=\pm1$ (front/back) the measurement effects the change $\omega\rightarrow\bm{p}_{x\pm}\in\Delta$, where
\begin{eqnarray}
\begin{aligned}
\bm{p}_{x+}\,:\ \ P\big(\omega=(+1,y,z)\big)=\tfrac{1}{4},\ \ P\big(\omega=(-1,y,z)\big)=0,\\
\bm{p}_{x-}\,:\ \ P\big(\omega=(+1,y,z)\big)=0,\ \ P\big(\omega=(-1,y,z)\big)=\tfrac{1}{4}.
\end{aligned}
\end{eqnarray}
Similarly, upon measuring $\mathbf{Y}=\pm1$ (right/left) the system gets disturbed as follows $\omega\rightarrow\bm{p}_{y\pm}\in\Delta$, where
\begin{eqnarray}
\begin{aligned}
\bm{p}_{y+}\,:\ P\big(\omega=(x,+1,z)\big)=\tfrac{1}{4},\ \ P\big(\omega=(x,-1,z)\big)=0,\\
\bm{p}_{y-}\,:\ P\big(\omega=(x,+1,z)\big)=0,\ \ P\big(\omega=(x,-1,z)\big)=\tfrac{1}{4}.
\end{aligned}
\end{eqnarray}
Finally, in result of measuring $\mathbf{Z}=\pm1$ (up/down) we are left with $\omega\rightarrow\bm{p}_{z\pm}\in\Delta$, where
\begin{eqnarray}
\begin{aligned}
\bm{p}_{z+}\,:\ \ P\big(\omega=(x,y,+1)\big)=\tfrac{1}{4},\ \ P\big(\omega=(x,y,-1)\big)=0,\\
\bm{p}_{z-}\,:\ \ P\big(\omega=(x,y,+1)\big)=0,\ \ P\big(\omega=(x,y,-1)\big)=\tfrac{1}{4}.
\end{aligned}
\end{eqnarray}
In short, the measurement reveals one of the coordinates and completely randomises the remaining ones. Note that since post-measurement states are also eigenstates of the respective measurement procedures, this definition guarantees \emph{repeatability} of measurements of the same kind. Due to measurement disturbance no further information gain about the initial state of the system is possible, i.e. the measurement is \emph{maximal}. As an aside we remark that so described system can be shown equivalent to a constrained version of a single qubit restricted to the convex hull of stabilizer states, Clifford transformations and Pauli observables; see~\cite{Bl13} for a detailed account.

\begin{figure*}[t]
\begin{center}
\includegraphics[width=\textwidth]{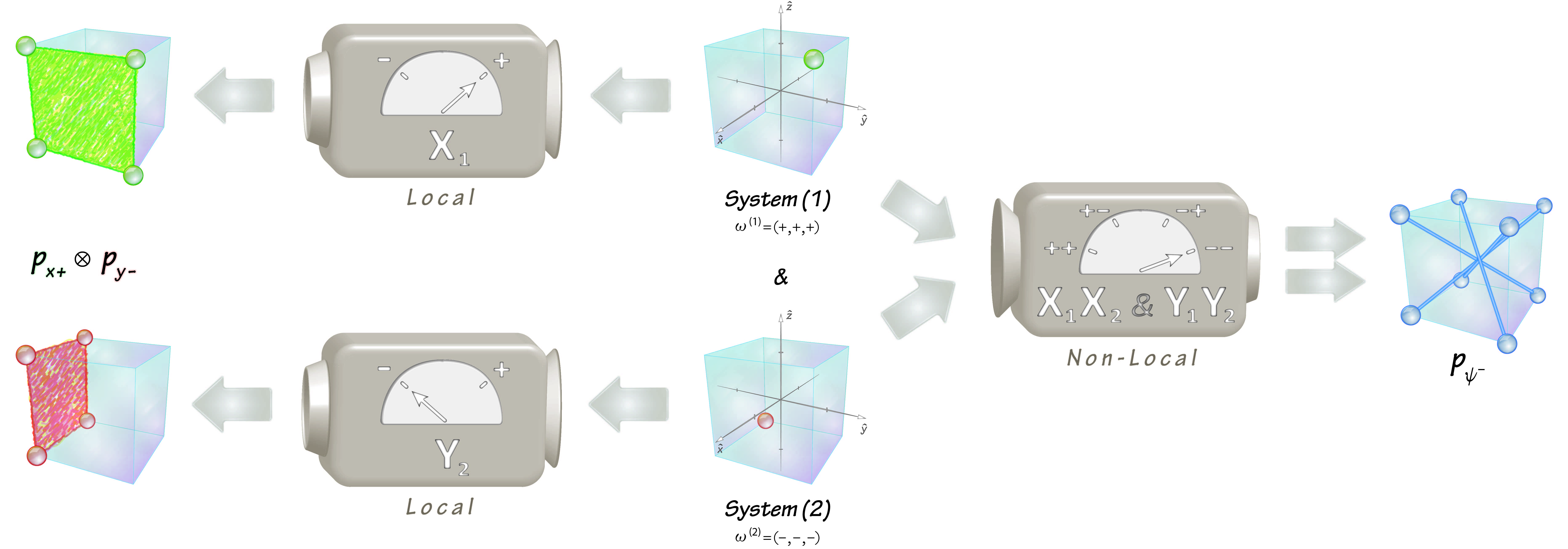}
\end{center}\caption{\label{Fig3} \textbf{Local vs. non-local measurement.} Two types of measurement procedures performed on the bipartite system (in the middle). On the left, measurement $\mathbf{X}_1\,\&\,\mathbf{Y}_2$ which consists in simultaneous local measurements of observables $\mathbf{X}_1$ and $\mathbf{Y}_2$ reveals that system $(1)$ was on the front ($x_1=+1$) and system $(2)$ on the left ($y_2=-1$) face of the respective cube, and leaves the bipartite system in the \emph{uncorrelated} product state $\bm{p}_{x+}\otimes\,\bm{p}_{y-}$. On the right, both elementary systems are brought together and undergo non-local measurement $\textbf{X}_1\textbf{X}_2\,\&\,\textbf{Y}_1\textbf{Y}_2$ which reveals joint information, described by observables $\mathbf{X}_1\mathbf{X}_2$ and $\mathbf{Y}_1\mathbf{Y}_2$, that systems occupied opposite faces in the $\hat{x}$ and $\hat{y}$ directions (i.e. $x_1x_2=-1$ and $y_1y_2=-1$) and subsequently leaves the bipartite system in the \emph{correlated} state $\bm{p}_{\psi^-}$.}
\end{figure*}

For the bipartite system above defined measurements may be performed independently on each component. Since these procedures do not affect the other (possibly distant) system we call them \emph{local measurements}. In total, we have $3\times3=9$ possible local measurement arrangements that will be denoted by $\mathbf{A}_1\,\&\,\mathbf{B}_2$, where $\mathbf{A},\mathbf{B}=\mathbf{X},\mathbf{Y},\mathbf{Z}$. Each such measurement reveals \emph{two bits} of information $(a_1,b_2)$ about the joint ontic state of the system $\omega=(\omega^{(1)},\omega^{(2)})$, where $\omega^{(i)}=(x_i,y_i,z_i)$ are the ontic states of the individual subsystems $(i=1,2)$, and leaves the joint system in the \emph{uncorrelated} product state $\omega\rightarrow \bm{p}_{a_1}\otimes\, \bm{p}_{b_2}\in \Delta^{(1)}\,\otimes\,\Delta^{(2)}$. We note that each state $\bm{p}_{a_1}\otimes\, \bm{p}_{b_2}$ is equiprobable mixture of $4\times4=16$ ontic states in $\Omega^{(1\,\&\,2)}$ which are compatible with the measurement result. See Fig.~\ref{Fig3} (on the left) for schematic illustration.
Observe that due to disturbance no further information can be inferred about the system, and hence this sort of measurement can be thought of as \emph{maximal} in this restricted setting. In general, one may choose to perform one of the $3\times3=9$ nontrivial local measurements $\mathbf{A}_1\,\&\,\mathbf{B}_2$ which provide information about two observables $\mathbf{A}_1$ and $\mathbf{B}_2$. Clearly, measurements of the same kind performed one after another are \emph{repeatable}. We note that it is entirely appropriate to think of such a measurement as two independent measurements made on each component separately.

\subsubsection{Non-local measurements}

Now, we proceed to define measurements which test correlations between components of the bipartite system. They will be called \emph{non-local measurements} since a straightforward realisation requires both systems brought together.

\begin{figure*}[t]
\begin{center}\includegraphics[width=0.93\textwidth]{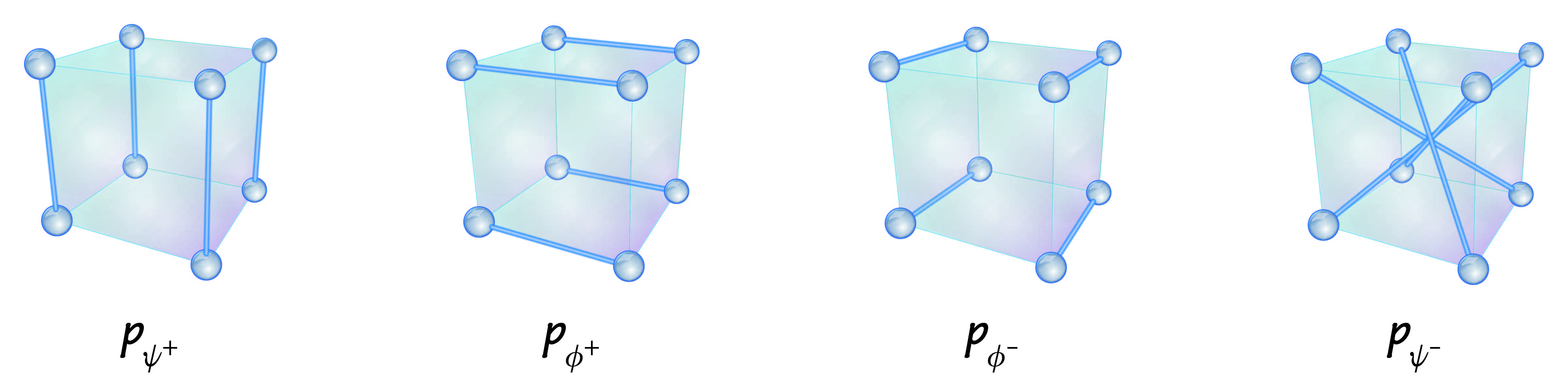}\vspace{-0.1cm} \\\hspace{0.5cm}\hrulefill\ \ \ \ \ \ \ \\\vspace{0.25cm}
\includegraphics[width=0.93\textwidth]{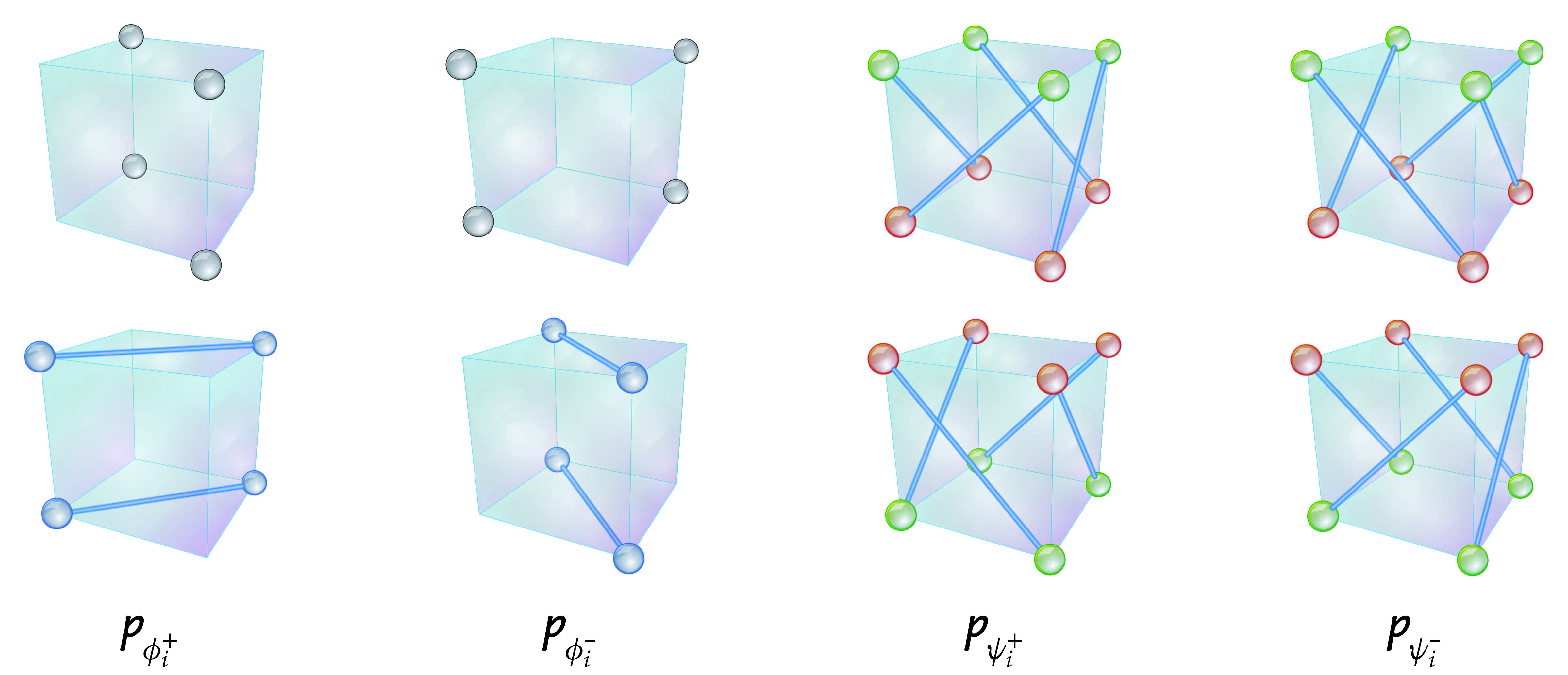}
\end{center}\caption{\label{Fig4} \textbf{Correlations in post-measurement states.} At the top, illustration of states  $\bm{p}_{\phi^\pm}$, $\bm{p}_{\psi^\pm}$ being equiprobable mixtures of eight (ontic) states, see Eq.~(\ref{psi-phi}). Each cube represents both systems at the same time with the following pictorial convention defining the set of allowed joint ontic  states: systems $(1)$ and $(2)$ may occupy only vertices at the opposite ends of the solid lines drawn on the pictures. Note that individually each system can take any of the eight vertices, but their joint position is highly \emph{correlated} (e.g. for state $\bm{p}_{\psi^-}$ the systems are bound to occupy opposite corners of the cubes). At the bottom, illustration of states $\bm{p}_{\phi_i^\pm}$, $\bm{p}_{\psi_i^\pm}$. Here, again,  states are equiprobable mixtures of eight (ontic) states of the joint systemsee Eq.~(\ref{psi-phi-i}). For states $\bm{p}_{\phi_i^\pm}$ the systems either occupy the same vertex chosen from the four vertices depicted in the upper cubes, or occupy opposite ends of the solid lines drawn on the cubes below. If we depict system $(1)$ as green and system $(2)$ as red, then for states $\bm{p}_{\psi_i^\pm}$ the systems may occupy only the respective ends of the solid lines drawn on the cubes. Like before, individually systems can take any of the eight vertices, but their joint position is highly \emph{correlated}.}
\end{figure*}

Let us imagine a device with two knobs which can be set to $\mathbf{A}_1\mathbf{B}_2\,\&\,\mathbf{C}_1\mathbf{D}_2$, where the choice of $\mathbf{A},\mathbf{B},\mathbf{C},\mathbf{D}$ ranges over $\mathbf{X},\mathbf{Y},\mathbf{Z}$; see Fig.~\ref{Fig3} (on the right). We assume that measurement $\mathbf{A}_1\mathbf{B}_2\,\&\,\mathbf{C}_1\mathbf{D}_2$ reveals \emph{two bits} of information $(a_1b_2,c_1d_2)$ about the joint (ontic) state of the system $\omega=(\omega^{(1)},\omega^{(2)})$, with $\omega^{(i)}=(x_i,y_i,z_i)$. For example, the measurement $\textbf{X}_1\textbf{X}_2\,\&\,\textbf{Y}_1\textbf{Y}_2$ tests the joint property of the bipartite system answering the following two questions: (i) are both systems together on the front or back face of the cube ($x_1x_2=\pm1$), and (ii) are both systems together on the right or left face of the cube ($y_1y_2=\pm1$); e.g. the result $(+1,-1)$ means that both systems occupy the same face as regards the $\hat{x}$ direction and different faces in the $\hat{y}$ direction.

In order to complete description of the measurement procedure we need to specify how it subsequently affects the system. We will assume that the measurement effects the change $\omega\rightarrow\bm{p}\in\Delta^{(1)}\otimes\Delta^{(2)}$ leaving the system in equiprobable mixture of eight appropriately chosen ontic states. For sake of simplicity in the following we will restrict ourselves only to $3+3=6$ non-local measurement settings listed in \textsl{Table}~1 and \textsl{Table}~2 (Fig.~\ref{C3-R3}).

Accordingly, for measurement $\textbf{X}_1\textbf{X}_2\,\&\,\textbf{Y}_1\textbf{Y}_2$ depending on the outcome, i.e. $(+1,+1)$, $(+1,-1)$, $(-1,+1)$, $(-1,-1)$, let the resulting post-measurement state be respectively:
\begin{eqnarray}\label{psi-phi}
\begin{aligned}
\bm{p}_{\psi^+}: \ \ P(\omega)=\tfrac{1}{8}\ \ \ \text{if}\ \ \ \ x_1=x_2\ ,\ \ y_1= y_2\ ,\ \ z_1\neq z_2\ ,\\
\bm{p}_{\phi^+}: \ \ P(\omega)=\tfrac{1}{8}\ \ \ \text{if}\ \ \ \ x_1=x_2\ ,\ \ y_1\neq y_2\ ,\ \ z_1=z_2\ ,\\
\bm{p}_{\phi^-}: \ \ P(\omega)=\tfrac{1}{8}\ \ \ \text{if}\ \ \ \ x_1\neq x_2\ ,\ \ y_1= y_2\ ,\ \ z_1=z_2\ ,\\
\bm{p}_{\psi^-}: \ \ P(\omega)=\tfrac{1}{8}\ \ \ \text{if}\ \ \ \ x_1\neq x_2\ ,\ \ y_1\neq y_2\ ,\ \ z_1\neq z_2\ ,
\end{aligned}\ 
\end{eqnarray}
and $P(\omega)=0$ otherwise. The pattern of correlations behind these definitions is best illustrated and analysed on pictures, see Fig.~\ref{Fig4} (at the top). In particular, one readily verifies that such defined measurement procedure is \emph{repeatable} (with states $\bm{p}_{\psi^\pm}$, $\bm{p}_{\phi^\pm}$ being the respective eigenstates) and \emph{maximal} (i.e. post-measurement disturbance prevents further information gain about the system). For measurements $\textbf{X}_1\textbf{X}_2\,\&\,\textbf{Z}_1\textbf{Z}_2$ and $\textbf{Z}_1\textbf{Z}_2\,\&\,\textbf{Y}_1\textbf{Y}_2$ let the assignment of post-measurement states to measurement outcomes follow the pattern given in \textsl{Table}~1 (Fig.~\ref{C3-R3}).

\begin{figure}[t]\small
\begin{center}\def\mc{}\renewcommand{\arraystretch}{1.8}
\begin{tabular}{c|cccc}
\textbf{\textsl{Table 1}}&$\ \ (+,+)$\ \ \ &$(+,-)$\ \ \ &$(-,+)$\ \ \ &$(-,-)$\ \ \ \\\hline\hline
$\textbf{X}_1\textbf{X}_2$\ \&\  $\textbf{Y}_1\textbf{Y}_2$ &\ $\bm{p}_{\psi^+}$ &  $\bm{p}_{\phi^+}$ & $\bm{p}_{\phi^-}$ & $\bm{p}_{\psi^-}$\\
$\textbf{X}_1\textbf{X}_2$\ \&\  $\textbf{Z}_1\textbf{Z}_2$ &\ $\bm{p}_{\phi^+}$ &  $\bm{p}_{\psi^+}$ & $\bm{p}_{\phi^-}$ & $\bm{p}_{\psi^-}$\\
$\textbf{Z}_1\textbf{Z}_2$\ \&\  $\textbf{Y}_1\textbf{Y}_2$ &\ $\bm{p}_{\phi^-}$ & $\bm{p}_{\phi^+}$  & $\bm{p}_{\psi^+}$ & $\bm{p}_{\psi^-}$\\
\end{tabular}
\\\vspace{0.5cm}
\begin{tabular}{c|cccc}
\textbf{\textsl{Table 2}}&$\ \ (+,+)$\ \ \ &$(+,-)$\ \ \ &$(-,+)$\ \ \ &$(-,-)$\ \ \ \\\hline\hline
$\textbf{X}_1\textbf{Y}_2$\ \&\  $\textbf{Y}_1\textbf{X}_2$ &\ $\bm{p}_{\phi_i^+}$ &  $\bm{p}_{\psi_i^-}$ & $\bm{p}_{\psi_i^+}$ & $\bm{p}_{\phi_i^-}$\\
$\textbf{Y}_1\textbf{X}_2$\ \&\  $\textbf{Z}_1\textbf{Z}_2$ &\ $\bm{p}_{\phi_i^+}$ &  $\bm{p}_{\psi_i^+}$ & $\bm{p}_{\phi_i^-}$ & $\bm{p}_{\psi_i^-}$\\
$\textbf{X}_1\textbf{Y}_2$\ \&\  $\textbf{Z}_1\textbf{Z}_2$ &\ $\bm{p}_{\phi_i^+}$ &  $\bm{p}_{\psi_i^-}$ & $\bm{p}_{\phi_i^-}$ & $\bm{p}_{\psi_i^+}$\\
\end{tabular}
\end{center}
\caption{\textbf{Assignment of post-measurement states.} Tables specify assignment of post-measurement states to the respective outcomes $(\pm1,\pm1)$ in non-local measurements considered in the model (left column). Measurements in each \textsl{Table}~1 and \textsl{Table}~2 share a common set of post-measurement states $\big\{\,\bm{p}_{\psi^\pm}\,,\,\bm{p}_{\phi^\pm}\big\}$ and $\big\{\,\bm{p}_{\psi_i^\pm}\,,\,\bm{p}_{\phi_i^\pm}\big\}$ respectively (for definitions of states see Eqs.~(\ref{psi-phi}) and (\ref{psi-phi-i})). As discussed in Sect.~\ref{Compatible-observables-and-sequential-measurements} these measurements performed in a sequence allow for simultaneous measurement of two sets of compatible observables: $\textbf{X}_1\textbf{X}_2$, $\textbf{Y}_1\textbf{Y}_2$, $\textbf{Z}_1\textbf{Z}_2$ (\textsl{Table}~1) and $\textbf{X}_1\textbf{Y}_2$, $\textbf{Y}_1\textbf{X}_2$, $\textbf{Z}_1\textbf{Z}_2$ (\textsl{Table}~2).}
\label{C3-R3}
\end{figure}

In a similar manner we define another three non-local measurements $\textbf{X}_1\textbf{Y}_2\,\&\,\textbf{Y}_1\textbf{X}_2$, $\textbf{Y}_1\textbf{X}_2\,\&\,\textbf{Z}_1\textbf{Z}_2$ and $\textbf{X}_1\textbf{Y}_2\,\&\,\textbf{Z}_1\textbf{Z}_2$ for which assignment of post-measurement states is listed in \textsl{Table}~2 (Fig.~\ref{C3-R3}). Note that in this case we use another set of epistemic states $\bm{p}_{\psi_i^\pm}$, $\bm{p}_{\phi_i^\pm}$ defined as follows:
\begin{eqnarray}\label{psi-phi-i}
\begin{aligned}
\bm{p}_{\phi_i^+}: \ \ P(\omega)=\tfrac{1}{8}\ \ \ \text{if}\ \ \ \ x_1=y_2\ ,\ \ y_1=x_2\ ,\ \ z_1=z_2
\ ,\\
\bm{p}_{\psi_i^-}: \ \ P(\omega)=\tfrac{1}{8}\ \ \ \text{if}\ \ \ \ x_1=y_2\ ,\ \ y_1\neq x_2\ ,\ \ z_1\neq z_2\ ,\\
\bm{p}_{\psi_i^+}: \ \ P(\omega)=\tfrac{1}{8}\ \ \ \text{if}\ \ \ \ x_1\neq y_2\ ,\ \ y_1=x_2\ ,\ \ z_1\neq z_2\ ,\\
\bm{p}_{\phi_i^-}: \ \ P(\omega)=\tfrac{1}{8}\ \ \ \text{if}\ \ \ \ x_1\neq y_2\ ,\ \ y_1\neq x_2\ ,\ \ z_1=z_2\ ,
\end{aligned}\ 
\end{eqnarray}
and $P(\omega)=0$ otherwise. See Fig.~\ref{Fig4} (at the bottom) for graphical illustration. A quick check shows that all measurements defined above are \emph{repeatable} and \emph{maximal} (with $\bm{p}_{\psi_i^\pm}$, $\bm{p}_{\phi_i^\pm}$ being the respective eigenstates). 

Note difference in the pattern defining states in Eqs.~(\ref{psi-phi}) and (\ref{psi-phi-i}). States $\bm{p}_{\psi^\pm}$, $\bm{p}_{\phi^\pm}$ are defined by even number of equalities ($2$ or $0$) and odd number of inequalities ($1$ or $3$), while in specification of states $\bm{p}_{\psi_i^\pm}$, $\bm{p}_{\phi_i^\pm}$ the number of equalities is odd ($3$ or $1$) and the number of inequalities is even ($0$ or $2$). Anticipating the analysis in Sect.~\ref{Contextuality-and-recovery-of-the-Mermin-Peres-square} we mention that this observation will be essential for reconstruction of the algebraic constraint in sequential measurements corresponding to column $\mathcal{C}_3$ and row $\mathcal{R}_3$ in the Mermin-Peres square.

This completes description of the choice of possible measurement procedures available in the model. Now, we proceed to discussion of observables and simultaneous measurability to show contextuality of the model and recover the pattern captured in the Mermin-Peres square.

\subsection{Analysis of the model}

The primary role of measurement is to reveal information about some property of the system which is called an observable. The scope of measurements available in our model is however very limited. In a measurement the agent can infer only \emph{partial information} about the system and further information gain is affected by \emph{post-measurement disturbance}.
These two restrictions built into the measurement process form the so called \emph{epistemic constraints} under which the agent operates investigating the system (see e.g.:~\cite{Sp07,BaRuSp12,Bl13}). In the following we discuss how these constraints affect information collected by the agent in a series of measurements.

\subsubsection{Observables and single-shot measurements}
\label{Observables-and-single-shot-measurements}

The model explicitly defines basic \emph{dichotomic} observables of the form $\textbf{A}_1$, $\textbf{B}_2$ and $\textbf{A}_1\textbf{B}_2$ where $\textbf{A},\textbf{B}=\textbf{X},\textbf{Y}\ \text{or}\ \textbf{Z}$. However, the measurements are so designed that the \emph{maximal} information gain about the bipartite system is restricted only to \emph{two bits}. If we choose to perform one of the local measurements $\textbf{A}_1\,\&\,\textbf{B}_2$, then the result $(a_1,b_2)$ provides direct access to two observables $\textbf{A}_1$ and $\textbf{B}_2$ with the respective values $a_1$ and $b_2$. Similarly by performing one of the non-local measurements $\textbf{A}_1\textbf{B}_2\,\&\,\textbf{C}_1\textbf{D}_2$ listed in \textsl{Table} 1 and \textsl{Table} 2 (Fig.~\ref{C3-R3}), the result $(a_1b_2,c_1d_2)$ gives direct access to two observables $\textbf{A}_1\textbf{B}_2$ and $\textbf{C}_1\textbf{D}_2$ whose values are respectively $a_1b_2$ and $c_1d_2$.

We can go further and observe that the information gained in a single-shot measurement can be used to calculate any function of observables that are being directly measured -- it is defined as the function of obtained results. This extends the scope of experimental procedures in which some observables can be measured. For example, in this way we get access to observables of the product type $\textbf{A}_1\textbf{B}_2$ also via the local procedure procedure $\textbf{A}_1\,\&\,\textbf{B}_2$ by simply multiplying the respective outcomes which provide the required result $a_1 b_2\equiv a_1\cdot b_2$. Hence in a single-shot local measurement $\textbf{A}_1\,\&\,\textbf{B}_2$ one can learn the values of three observables $\textbf{A}_1$, $\textbf{B}_2$ and $\textbf{A}_1\textbf{B}_2$ at once. 
We conclude that the model admits various experimental procedures for measuring product observables $\textbf{A}_1\textbf{B}_2$, i.e. one can either measure them directly in one of the non-local measurements $\textbf{A}_1\textbf{B}_2\,\&\,\textbf{C}_1\textbf{D}_2$ or choose to perform local measurement $\textbf{A}_1\,\&\,\textbf{B}_2$ and calculate the result as explained above. 

Let us observe that single-shot measurements reveal 'true' information about the system, i.e. values of the associated observables are calculated according to the actual values of coordinates assumed by the ontic state of the system. We will call such a measurement \emph{faithful}. Then, after the measurement the system gets disturbed which prevents further gain of information about its initial ontic state. This means that although subsequent measurement is faithful as concerns the state of the system after the first measurement, in general it is not true to the initial situation before the first measurement. Note, however, that the nature of state disturbance defined in the model is such that after the measurement the system may assume only those ontic states which are compatible with the obtained outcome. This entails that the value assigned to the chosen basic observable $\textbf{A}_1$, $\textbf{B}_2$ and $\textbf{A}_1\textbf{B}_2$ is \emph{repeatable}, i.e. subsequent measurement of the same observable repeats the  result (even if the measurements are of different kind).

\subsubsection{Compatible observables and sequential measurements}
\label{Compatible-observables-and-sequential-measurements}

Now let us consider a collection of dichotomic observables. Each of them can be measured individually in a single-shot experiment as explained in the preceding Sect.~\ref{Observables-and-single-shot-measurements}. Now, suppose we make a \emph{sequential measurement} which consists of a series of such measurements performed one after another on the same system. As a result we get a series of outcomes which associate values to observables of interest. In general, such a sequence of measurements will not be repeatable in a sense that various instances of measurements in the sequence testing the same observable will produce different results. However, sometimes it may happen that for some observables appropriate choice of measurements will make the results for the same observables repeat. In more precise terms, this means that for a given collection of observables one may find a collection of single-shot measurements testing these observables with the following property: in any sequential measurement, no matter of its length and order, the outcomes associated to these observables in consecutive measurements are \emph{repeatable} (this should be true for any conceivable initial state of the system). Then such a collection of observables is called \emph{compatible} and according to common practice considered to be \emph{simultaneously measurable}. Values assigned to compatible observables in such a measurement are the respective outcomes obtained in the sequential measurement (due to repeatability the assignment is consistent).

It is important to note that the definition of simultaneous measurability is based on the paradigm of repeatability and does \emph{not} refer to faithfulness which requires knowledge of the 'true' values of the underlying ontic state of the system. This kind of approach is characteristic for operational theories. It is very natural in the context of our model, since the agent subject to epistemic constraints (hence without direct access to the ontic state) needs to judge only by himself what constitutes an observable and how it is measured. Here, prescription is simple: for simultaneous measurability of a collection of observables it is enough to point out a collection of measurements which performed in a sequential manner always give repeatable outcomes. 

For illustration of the concept of of compatibility we give a few simple examples. Collection of two observables directly measurable in one of the local/non-local measurements allowed in the model is a trivial example of compatible set; this is because a sequence of the same measurements is always repeatable. By the same token we infer that any collection of three observables $\textbf{A}_1$, $\textbf{B}_2$ and $\textbf{A}_1\textbf{B}_2$ is compatible, since from the discussion in Sect.~\ref{Observables-and-single-shot-measurements} we know that it can be measured in a single-shot local-measurement $\textbf{A}_1\,\&\,\textbf{B}_2$ (with post-measurement states $\bm{p}_{a\pm}\otimes\,\bm{p}_{b\pm}$). 

In order to give a more sophisticated example we recall that repeatability of measurements defined in the model is guaranteed by a specific definition of post-measurement states which are also their eigenstates, i.e. subsequent measurement on the system in such a state will necessarily reproduce the result and the state. It follows that a sufficient condition for repeatability of results in a sequence of possibly different measurements is that they share the same set of post-measurement states. In such a case, the first measurement chooses the post-measurement state which, being eigenstate of all the measurements, remains fixed throughout the whole sequence of measurements.

Now we are in position to give two nontrivial examples of compatible sets in the model. The first one is a collection of three observables $\textbf{X}_1\textbf{X}_2$, $\textbf{Y}_1\textbf{Y}_2$ and $\textbf{Z}_1\textbf{Z}_2$. These observables can be measured in a sequence of non-local measurements listed in \textsl{Table} 1 (Fig.~\ref{C3-R3}). Recall that all these measurement procedures share the same set of post measurement states $\big\{\,\bm{p}_{\psi^\pm}\,,\,\bm{p}_{\phi^\pm}\big\}$. This entails repeatability of results in any sequence of such measurements, and hence compatibility of the considered collection of observables. In a smiler manner one argues that a collection of three observables $\textbf{X}_1\textbf{Y}_2$, $\textbf{Y}_1\textbf{X}_2$ and $\textbf{Z}_1\textbf{Z}_2$ is compatible. In this case, we choose to perform a sequence of non-local measurements taken from \textsl{Table} 2 (Fig.~\ref{C3-R3}) with the common set of post-measured states $\big\{\,\bm{p}_{\psi_i^\pm}\,,\,\bm{p}_{\phi_i^\pm}\big\}$.

\subsubsection{Contextuality and recovery of the Mermin-Peres square}
\label{Contextuality-and-recovery-of-the-Mermin-Peres-square}

\begin{figure*}[t]
\begin{center}
\includegraphics[width=1\textwidth]{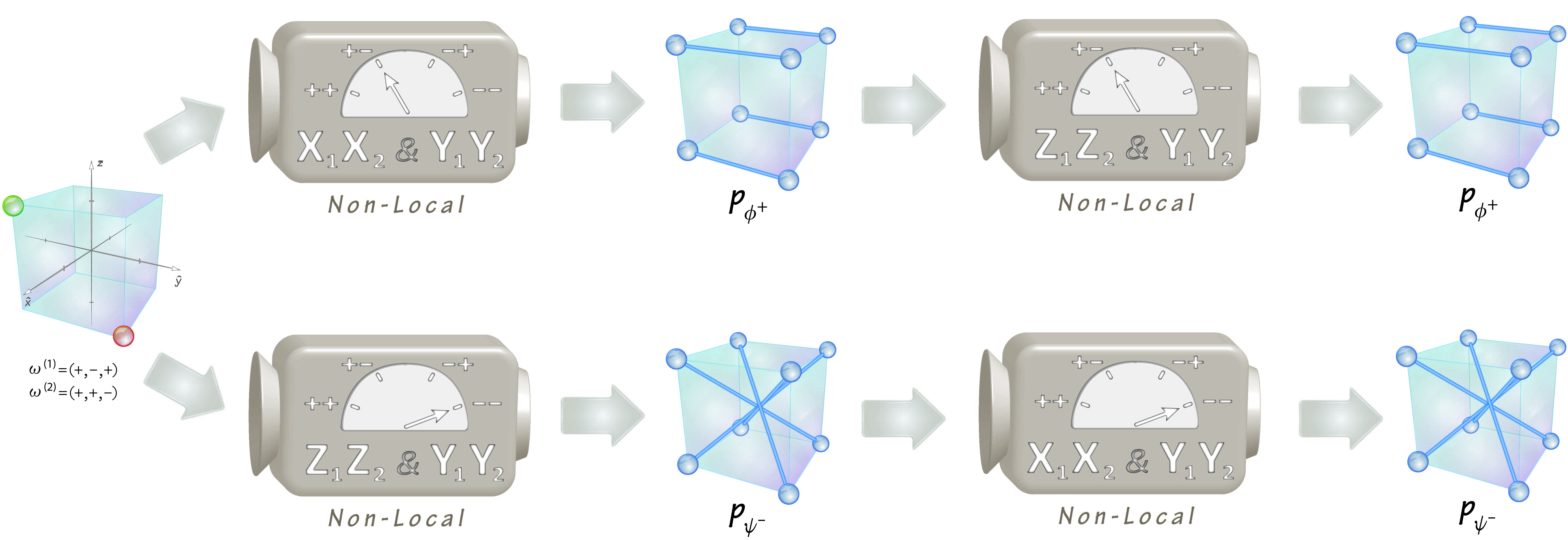}
\end{center}\caption{\label{Fig7} \textbf{Context in sequential measurements.} Two sequential measurements of three compatible observables  $\textbf{X}_1\textbf{X}_2$, $\textbf{Y}_1\textbf{Y}_2$ and $\textbf{Z}_1\textbf{Z}_2$ performed in different order on the bipartite system in the same initial ontic state $\omega=(\omega^{(1)},\omega^{(1)})$, where $\omega^{(1)}=(+1,-1,+1)$ (depicted as green) and $\omega^{(2)}=(+1,+1,-1)$ (depicted as red). Although outcomes in each sequence are repeatable, the values assigned to measured observables are \emph{different}. At the top the values read $\textbf{X}_1\textbf{X}_2=+1$, $\textbf{Y}_1\textbf{Y}_2=-1$ and $\textbf{Z}_1\textbf{Z}_2=+1$, while at bottom for the reversed order of experiments they are  $\textbf{X}_1\textbf{X}_2=-1$, $\textbf{Y}_1\textbf{Y}_2=-1$ and $\textbf{Z}_1\textbf{Z}_2=-1$. This discrepancy stems from the fact that only the first measurement provides the 'true' values of the initial ontic state of the system and disturbs the system accordingly. The subsequent measurements, however, reveal information only about so prepared post-measurement state. According to \textsl{Table}~1 (Fig.~\ref{C3-R3}) it is $\phi^+$ at the top, and $\psi^-$ at the bottom. In this sense, the first measurement provides a particular \emph{context} (identified with the post-measurement state) in which outcomes in the remaining measurements are already decided, but not have to be true to the initial ontic state of the system (in this example it is certainly not the case, since for all the values to be faithful to the initial state they would have to read $\textbf{X}_1\textbf{X}_2=+1$, $\textbf{Y}_1\textbf{Y}_2=-1$ and $\textbf{Z}_1\textbf{Z}_2=-1$).}
\end{figure*}

Let us begin with an important remark that information about the system revealed in a sequential measurement of a compatible set does not have to be faithful -- the only requirement for compatibility of a set of observables is repeatability. In our case, we observe that in a sequence of measurements only the first one is bound to disclose the actual value of the associated observables (like in the single-shot measurement), whereas the values revealed in the subsequent measurements of the remaining compatible observables can be affected by state disturbance (and hence no longer reflect the initial state of affairs). We have seen that in the model the maximal 'true' information that can be learned about the system is limited only to two bits. In a sequential experiment upon the first measurement values of the remaining compatible observables, that are yet to be revealed, get altered (not necessarily in accord with the initial ontic state) and encoded in the common post-measurement state (cf. Fig.~\ref{Fig7}). For example, take the collection of three compatible observables $\textbf{X}_1\textbf{X}_2$, $\textbf{Y}_1\textbf{Y}_2$ and $\textbf{Z}_1\textbf{Z}_2$. We know that it can be measured in a sequence of at least two different non-local measurements from \textsl{Table}~1 (Fig.~\ref{C3-R3}). If we begin with measurement $\textbf{X}_1\textbf{X}_2\,\&\,\textbf{Y}_1\textbf{Y}_2$ and obtain the faithful outcome, say $(+,-)$, then post-measurement state $\bf{p}_{\phi^+}$ dictates outcomes for all subsequent measurements, e.g. measurement of  $\textbf{Z}_1\textbf{Z}_2\,\&\,\textbf{Y}_1\textbf{Y}_2$ will certainly yield outcome $(+,-)$ and again reproduce $\bf{p}_{\phi^+}$. In this case assignment of values to compatible observables $\textbf{X}_1\textbf{X}_2$, $\textbf{Y}_1\textbf{Y}_2$ and $\textbf{Z}_1\textbf{Z}_2$ reads $+1$, $-1$ and $+1$ respectively. Observe that the value taken by $\textbf{Z}_1\textbf{Z}_2$ does not have to be faithful and might have been different if measurements were performed in different order. See Fig.~\ref{Fig7} for illustration.

Extension of the notion of simultaneous measurability to compatible observables requires further qualification of the measurement context when sequential measurements are taken into account. We have seen that
in a sequential measurement of a collection of compatible observables the first one plays a crucial role. It reveals 'true' values of the initial ontic state of the system and dictates in advance the values of observables to be assigned in the subsequent measurements. All this information is contained in the post-measurement state picked out by the first measurement. Essential for simultaneous measurability of a collection of compatible observables is the same set of post-measurement states shared by all measurements in the sequence. We say that this common set provides a \emph{context} in which compatible observables are probed.\footnote{Observe that we use the term simultaneous measurability as it is understood in quantum theory. That is, in spite the fact that the measurements can not be realised all at the same time, one relaxes the condition of simultaneity to include sequential measurements of a set of observables when they are repeatable. This boils down to the notion of compatible observables defined through the property of having a common eigenbasis which defines the \emph{context} in which observables are measured. In the quantum version of the Mermin-Peres square (Fig.~\ref{Quantum-Mermin-Peres-Square}) the respective bases are product states $|\pm x\rangle\otimes|\pm x\rangle$, $|\pm y\rangle\otimes|\pm y\rangle$, $|\pm x\rangle\otimes|\pm y\rangle$, $|\pm y\rangle\otimes|\pm x\rangle$  for $\mathcal{R}_1$, $\mathcal{R}_2$, $\mathcal{C}_1$, $\mathcal{C}_2$ respectively, and entangled states $|\psi^\pm\rangle$, $|\phi^\pm\rangle$ for $\mathcal{C}_3$ and $|\psi_i^\pm\rangle$, $|\phi_i^\pm\rangle$ for $\mathcal{R}_3$.} The first measurement picks one of them which in advance decides all the values that will be assigned in the subsequent measurements. Note that this assignment could have ben different had we chosen a different way of measuring given observable (cf. Fig.~\ref{Fig7}). We conclude that the model is highly \emph{contextual}. In the following we will show that the character of contextuality mimics the pattern of the Mermin-Peres square.

In Sect.~\ref{Compatible-observables-and-sequential-measurements} we have discussed a few sets of compatible observables in the model. For our purposes we shall consider only six of them that are listed below (together with description how to measure them):
\begin{eqnarray}\nonumber
\mathcal{R}_1\ :&\textbf{X}_1,\ \textbf{X}_2,\ \textbf{X}_1\textbf{X}_2&\Leftarrow\text{ single local measurement }\ \textbf{X}_1\,\&\,\textbf{X}_2\\\nonumber
\mathcal{R}_2\ :&\textbf{Y}_1,\ \textbf{Y}_2,\ \textbf{Y}_1\textbf{Y}_2&\Leftarrow\text{ single local measurement }\ \textbf{Y}_1\,\&\,\textbf{Y}_2\\\nonumber
\mathcal{R}_3\ :&\textbf{X}_1\textbf{Y}_2,\ \textbf{Y}_1\textbf{X}_2,\ \textbf{Z}_1\textbf{Z}_2&\Leftarrow\text{ any sequence of non-local measurement from \textsl{Table}~2}\\\nonumber
\mathcal{C}_1\ :&\textbf{X}_1,\ \textbf{Y}_2,\ \textbf{X}_1\textbf{Y}_2&\Leftarrow\text{ single local measurement }\ \textbf{X}_1\,\&\,\textbf{Y}_2\\\nonumber
\mathcal{C}_2\ :&\textbf{Y}_1,\ \textbf{X}_2,\ \textbf{Y}_1\textbf{X}_2&\Leftarrow\text{ single local measurement }\ \textbf{Y}_1\,\&\,\textbf{X}_2\\\nonumber
\mathcal{C}_3\ :&\textbf{X}_1\textbf{X}_2,\ \textbf{Y}_1\textbf{Y}_2,\ \textbf{Z}_1\textbf{Z}_2&\Leftarrow\text{ any sequence of non-local measurement from \textsl{Table}~1}
\end{eqnarray}
Note that they provide analogues of six compatible sets of dichotomic observes corresponding to rows and columns in the Mermin-Peres square, see Fig.~\ref{Quantum-Mermin-Peres-Square}. In order to reconstruct the pattern which leads to the contradiction with non-contextuality we need to check the products of values that are assigned to these observables in the respective measurements of these compatible sets.

First, let us consider observables in row $\mathcal{R}_1$ that are measured in a single-shot measurement $\textbf{X}_1\,\&\,\textbf{X}_2$ which provides outcome $(x_1,x_2)$. Values assigned to $\textbf{X}_1$, $\textbf{X}_2$, $\textbf{X}_1\textbf{X}_2$ are respectively $x_1$, $x_2$ and $x_1x_2\equiv x_1\cdot x_2$. Hence their product reads: $x_1\cdot x_2\cdot x_1x_2=(x_1)^2(x_2)^2=+1$. By the same token irrespective of the initial state one readily checks that the product of values assigned to compatible sets in rows $\mathcal{R}_1$, $\mathcal{R}_2$ and columns $\mathcal{C}_1$, $\mathcal{C}_2$ is always equal to $+1$.

Next, we verify column $\mathcal{C}_3$. We have argued that in the sequential measurement information about all the outcomes is already encoded in the first post-measurement state chosen from $\{\,\bf{p}_{\psi^\pm},\bf{p}_{\phi^\pm}\}$. Since it remains the same throughout the whole sequence of measurements \emph{all} values revealed by the subsequent measurements will be consistent with the definition of the given state in Eq.~(\ref{psi-phi}). Note that in each case the state is described by three conditions which consist of even number of equalities ($2$ or $0$) and odd number of inequalities ($1$ or $3$). This entails that independently of the initial state values assigned to observables in $\mathcal{C}_3$ will always satisfy: $x_1x_2\cdot y_1y_2\cdot z_1z_2\equiv (x_1\cdot x_2)\,(y_1\cdot y_2)\,(z_1\cdot z_2)=-1$ (since it is a product of odd number of $-1$'s and the rest $+1$'s).

In a similar manner we analyse row $\mathcal{R}_3$. Here, the common set of post-measurement states is $\{\,\bf{p}_{\psi_i^\pm},\bf{p}_{\phi_i^\pm}\}$. Each of the states in Eq.~(\ref{psi-phi-i}) is defined by there conditions with odd number of equalities ($3$ or $1$) and even number of inequalities ($0$ or $2$). Hence the product of values assigned to compatible observables in row $\mathcal{R}_3$ will always be: $x_1y_2\cdot y_1x_2\cdot z_1z_2\equiv (x_1\cdot y_2)\,(y_1\cdot x_2)\,(z_1\cdot z_2)=+1$ (since it is a product of even number of $-1$'s and the rest $+1$'s).

This concludes the check of properties required for the contradiction with non-contextuality in the Mermin-Peres square, i.e. (i) compatibility of observables along rows and columns in Fig.~\ref{Quantum-Mermin-Peres-Square}, and (ii) product of their values assigned in the associated measurements equal to $+1$ except for column $\mathcal{C}_3$ where the product is $-1$. In this way we confirmed our earlier observation that the model is contextual. Moreover, it shows the same character of contextuality as the quantum case in the Mermin-Peres square (cf. Sect.~\ref{MP-square}). In addition, from the above discussion of the model it follows that reconstruction of the algebraic pattern of observables does not depend on the state of the system nor the order of sequential measurements. In consequence, we get state-independent violation of the associated contextuality-inequalities~\cite{Ca08d} which is insensitive to the order of measurements (cf.~\cite{SzKlGu13}).

As a final remark, recall that for local measurements $\mathbf{A}_1\,\&\,\mathbf{B}_2$ the bipartite system is left in the uncorrelated mixture of $4\times4=16$ ontic states. Let us emphasise that the character of non-local measurement $\mathbf{A}_1\mathbf{B}_2\,\&\,\mathbf{C}_1\mathbf{D}_2$ is essentially different. On top of revealing joint information about both components it introduces probabilistic correlations between the components after the measurement as defined in Eqs.~(\ref{psi-phi}) and (\ref{psi-phi-i}). Moreover, the resulting post-measurement state is equiprobable mixture of only 8 ontic states -- it is twice less than the support of post-measurement state in the local measurement. This can be seen as the cost of introducing contextual effects into the model which should be compared with the discussion of memory cost of simulating contextuality by Mealy machines~\cite{KlGuPoLaCa11}. However, we note that the present model differs from the latter in two important aspects. Firstly, it is build upon different ontology which is intrinsically probabilistic and $\psi$-epistemic~\cite{Sp07,HaSp10}. Secondly, contextual effects in the model arise from probabilistic correlations in post-measurement states rather than external memory (i.e. additional information required for simulation of contextuality in the model is hidden solely in correlations).

\section{Discussion}
\label{Discussion}

In summary, we have presented classical model of the bipartite system and described two kinds of measurement procedures which take as a pattern the structure of quantum observables in the Mermin-Peres square. Observables discussed in the model have the same (non-)local character, are sensitive to the measurement context, preserve simultaneous measurability along rows and columns, as well as uphold the curious property held by the product of values. Given all this, the model reconstructs all features of the Mermin-Peres square relevant to the discussion of contextuality. Accordingly, quantum arguments translate verbatim to our case and hence the model contradicts the assumption of non-contextuality precisely in the same manner as quantum observables considered in the Mermin-Peres square. Note that in constructing the model we proceeded in close parallel to the operational framework of quantum mechanics. In effect, on top of simulating quantum contextuality, we get a classical analogue which reflects the structure and behaviour of a bipartite quantum system probed by sequential measurements in the restricted setting of the Mermin-Peres square. In particular, the model violates the associated contextuality-inequality~\cite{Ca08d} in the same manner as quantum mechanics does; the violation is state-independent and insensitive to the order of measurements.

This clearly demonstrates that classical systems can be contextual too, and indeed in a very  characteristic quantum-like fashion. Hence the conclusion that contextuality by itself is not enough to be taken as a signature of ''quantumness''. In fact, the model shows that contextual behaviour may be simply an effect of limited information gain and post-measurement state disturbance, both being a plausible scenario in the classical realm too. It can be classified as a contextual hidden variable model with the ontic states $\Omega^{(1\,\&\,2)}\equiv\Omega^{(1)}\times\,\Omega^{(2)}$ playing the role of hidden variables, and contextuality deriving from subtle distinction between the concept of simultaneous measurability, faithfulness, compatible observables and the actual sequential measurements providing the context. The model belongs to the category of $\psi$-epistemic models in the classification of~\cite{HaSp10}. Let us point out that our construction differs from the existing proposals which can be immediately observed from the structure of the ontic state space (cf.~\cite{La09a,KlGuPoLaCa11,La12,SzKlGu13,LeJeBaRu12,EmDeSuVe13}). Moreover, it presents a different ontology which in the model builds upon the classical picture of a bipartite system being composed of two separate components (with all effects resulting from classical correlations).

If contextuality by itself is not a token of non-classicality, then what makes quantum theory so different? Or more generally, which conceptual features distinguish quantum mechanics from classical theories. This sort of questions occupy a profound place in quantum foundations. Recently a considerable progress has been made in separating quantum from classical effects by means of toy models (see e.g.~\cite{Sp07,Ha99,Ki03,La09a,DaPlPl02,KlGuPoLaCa11,BaRuSp12,La12,SzKlGu13,Bl13}) and study of $\psi$-epistemic reconstructions.
In this paper, we aimed at demystifying the concept of contextuality by showing that it manifests in the classical regime too. The presented model contributes to an often debated topic whether or not contextuality is a 'true' signature of non-classicality. Clearly, the opinion that it is typically quantum effect is not fully justified and requires further qualification (e.g. by bringing separability and non-locality to the spotlight~\cite{Me93}).


\bibliographystyle{model1-num-names}
\bibliography{CombQuant}







\end{document}